# Controlled suppression of local magnetization hysteresis by dc in-plane field in $Bi_2Sr_2CaCu_2O_8$

I. Gutman[1] , S. Goldberg[1], Y. Myasoedov[1], E. Zeldov[1] and T. Tamegai[2]
*1 Department of Condensed Matter Physics, Weizmann Institute of Science, Rehovot 76100, Israel*
*2 Department of Applied Physics, The University of Tokyo, Hongo, Bunkyo-ku, Tokyo 113-8656, Japan*

The results about controlled suppression of local magnetic hysteresis in $Bi_2Sr_2CaCu_2O_8$ are presented. This suppression is obtained for temperature of 82 K and normal magnetic fields up to 20 Oe by applying a dc in-plane field up to 20 Oe. This work demonstrates that in the mentioned above region of phase diagram a dc in-plane field overcomes hysteresis mainly through the mechanism of geometrical barrier (GB). GB arises due to competition between the line energy of a vortex penetrating the sample and the Lorentz force of Meissner currents which focuses vortices in the sample's center to form a dome-shaped vortex distribution. In presence of in-plane field, we find vortex chains, formed by Josephson vortices, outside the dome that apparently extend all the way to the sample edges. We propose that these chains act as channels for vortex system equilibration. Quantitative energy estimations showed that vortex chain energy is high enough to move the dome edges resulting in magnetization hysteresis suppression.

## I. INTRODUCTION

Local magnetization hysteresis is among the most extensively studied features of high-temperature superconductors (HTSC). The usual source of hysteresis in superconductors is bulk vortex pinning due to material defects [1,2]. Two additional known mechanisms of magnetic irreversibility are the Bean-Livingston (BL) surface barrier [3,4] and the geometrical barrier (GB) [5-7].

GB arises due to a competition between the line energy of a vortex penetrating into the sample and the Lorentz force of Meissner currents which focuses vortices in the sample's center to form a dome-shaped vortex distribution. This work demonstrates that dc in-plane field overcomes hysteresis mainly through the GB suppression in a region of phase diagram under consideration (high temperatures and low fields).

## II. EXPERIMENTAL DETAILS

We have carried out magneto-optical (MO) measurements in Bi2Sr2CaCu2O8 crystal near the superconducting transition under crossing magnetic fields.

In standard MO measurements a magnetic field ($H_z$) is applied and the corresponding distribution of the local induction measured on the surface of the sample using a MO garnet indicator and a polarized light microscope. In order to investigate the effect of dc in-plane Field ($H_x$) with better resolution, we used modification of the standard MO measurements so that either the applied field, or the transport current through the sample are periodically modulated by a small amount and the corresponding differential signal is recorded by the CCD camera.

The difference between the two kinds of modulations is that field modulation affects vortex density making them breathing in 2D. In contrast, current modulation applies on the vortices a Lorentz force, which makes their 1D translation hoping to achieve a better resolution.

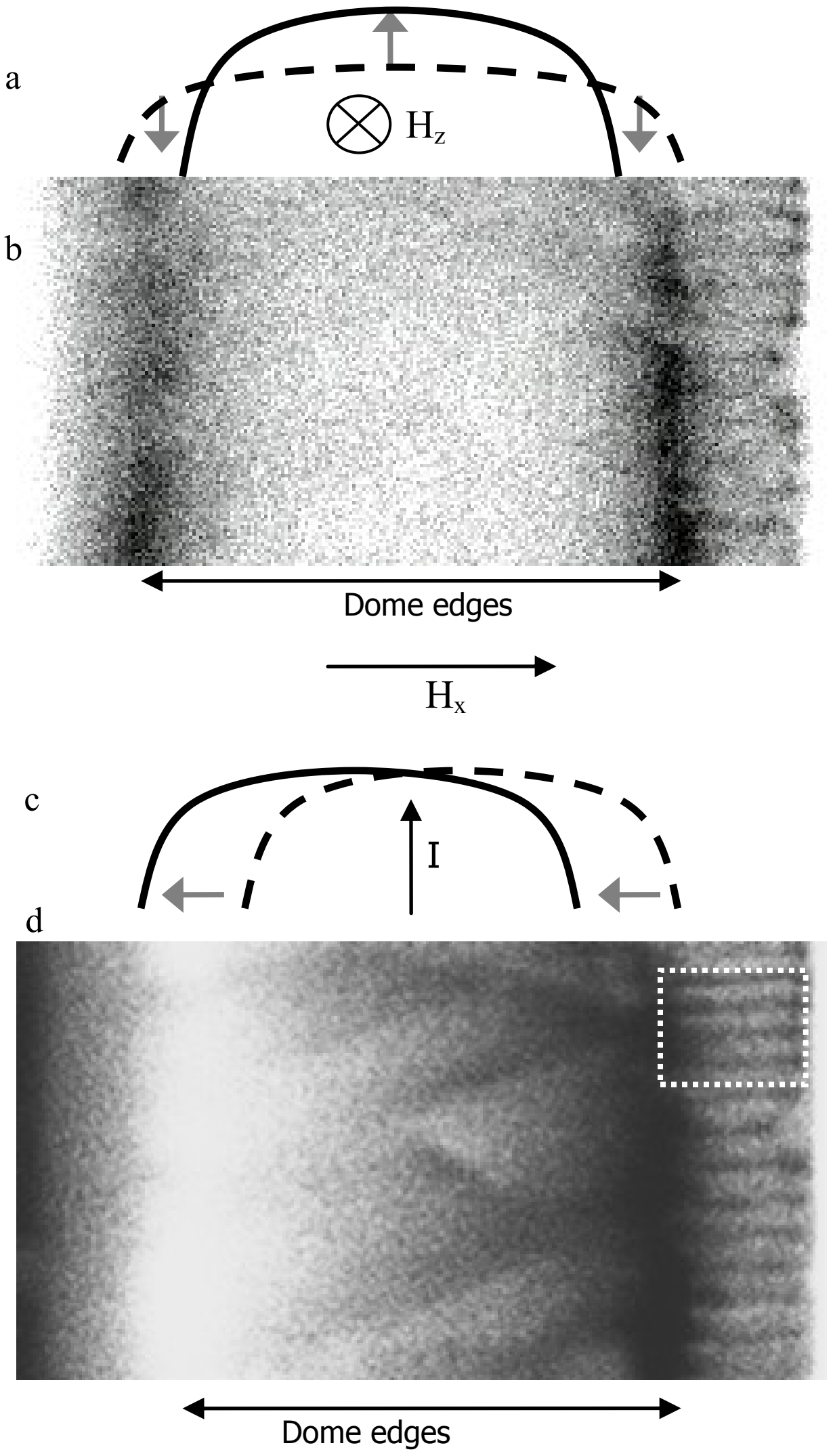

Figure 1: Differential MO images of $Bi_2Sr_2CaCu_2O_8$ crystal with modulation of b) applied field $dH_z$=1 Oe. d) applied current dI=60 mA. $H_x$=15 Oe , $H_z$=9.4 Oe T=82 K. a),c) provide schematic explanation to the images.

## III. RESULTS

The images in Fig. 1 present vortex dome profiles measured by two kinds of modulations: modulation with respect to the perpendicular magnetic field $H_Z$ (see Fig. 1b) and to the current I (see Fig. 1d).

Figure 1b is a modulation of $H_z$ by 1 Oe. The dome edges are visualized as minimas obtained by subtracting the reduced dome at $H_z$-$dH_z$/2 (dashed curve in Fig. 1a) from the expanded dome at $H_z$+$dH_z$/2 (continuous curve in Fig. 1a).

The image in Figure 1d is obtained by modulating the current by 60 mA. The dashed (+dI/2) and continuous (-dI/2) curves in Fig. 1c are two dome profiles that are shifted, in respect to each other, by the current. Their subtraction reproduces the experimental profile.

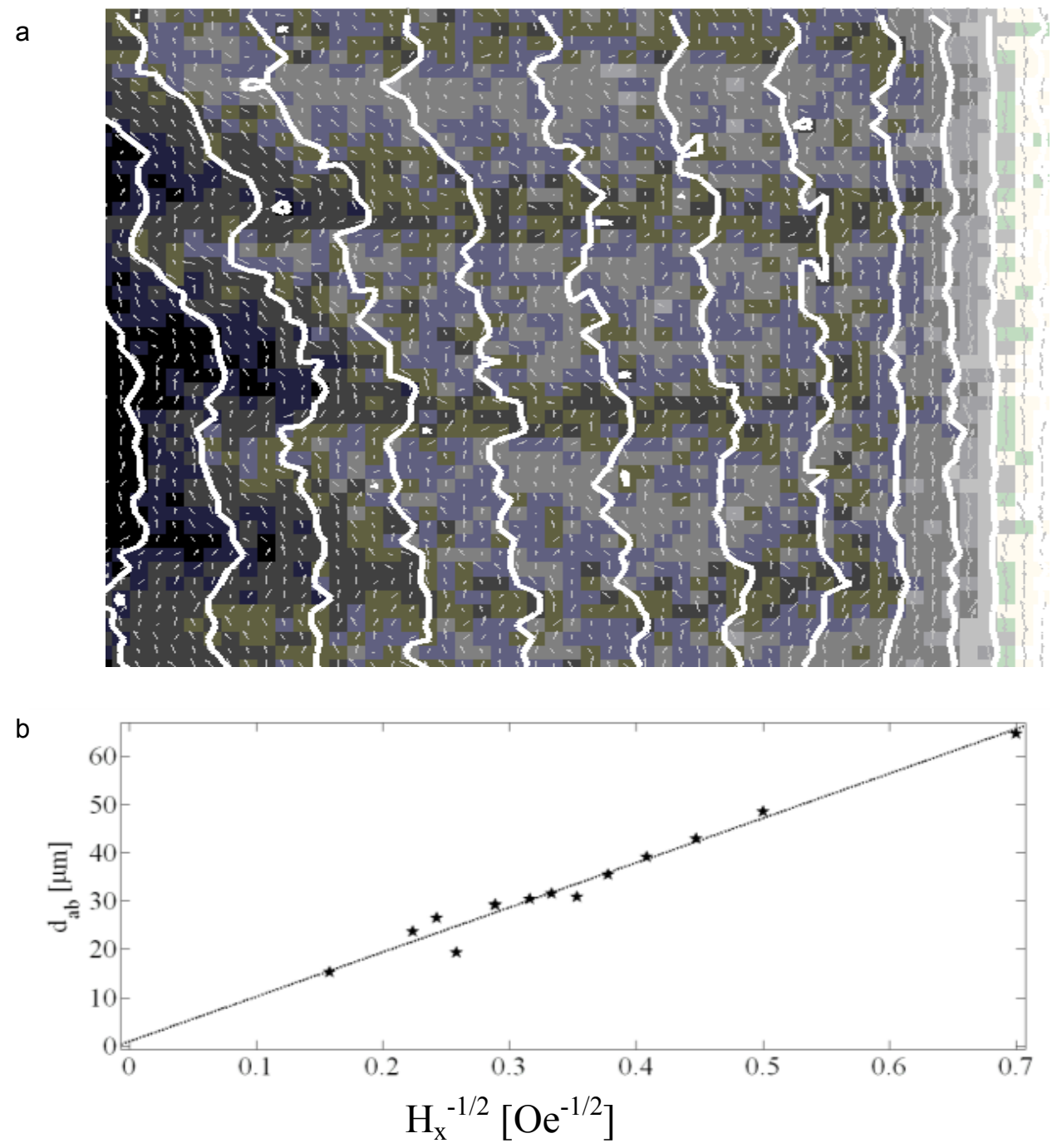

Figure 2: a) 2D current inversion method reveals the current direction across the chains of the dashed white square in Fig. 1d. b) Linear dependence of vortex separation with $H_x^{-1/2}$.

**A significant enhancement in the resolution in the case of current modulation is noticed.** This enhancement can be seen in the sharpness of the vortex dome edges and vortex chains seen as black horizontal periodic strips on the right side of each image.

In figure 2a a 2D current inversion procedure [8,9] was implemented to understand the current flow through the chains. This is an enlarged image of the white dashed rectangle in Fig. 1d. Current flow trajectories are the white curves and the current field is marked by gray arrows. Notice that the current "jumps" over the chains and tries to surround them.

Vortex chains extend with $H_z$, apparently all the way to the sample edges. The density of chains grows with $H_x$ and the scaling of the distance between the chains vs. $H_x$ is consistent with vortex chains formed by Josephsone vortices (JVs). Figure 2b show a linear dependence of vortex chain separation on $H_x^{-1/2}$ which is in agreement with theory of JV's which predicts

(1) $$d_{chains} = \sqrt{\sqrt{3}\phi_0\gamma / 2H_x}$$ .

The slope of the fit gives the anisotropy parameter $\gamma$=406 with a good agreement to other published estimates [10, 11].

Thus, the chains are present in the regions that are “vortex free” (outside the dome) in the absence of an in-plane field $H_x$ and are formed by pancake vortices (PVs) coupled to JVs.

Correlation between two kinds of hysteresis is shown in Fig. 3: hysteresis in local magnetization determined by B-$H_z$ and in dome edges location. Both types of hysteresis were evaluated at the point $H_z$=9.4 Oe.

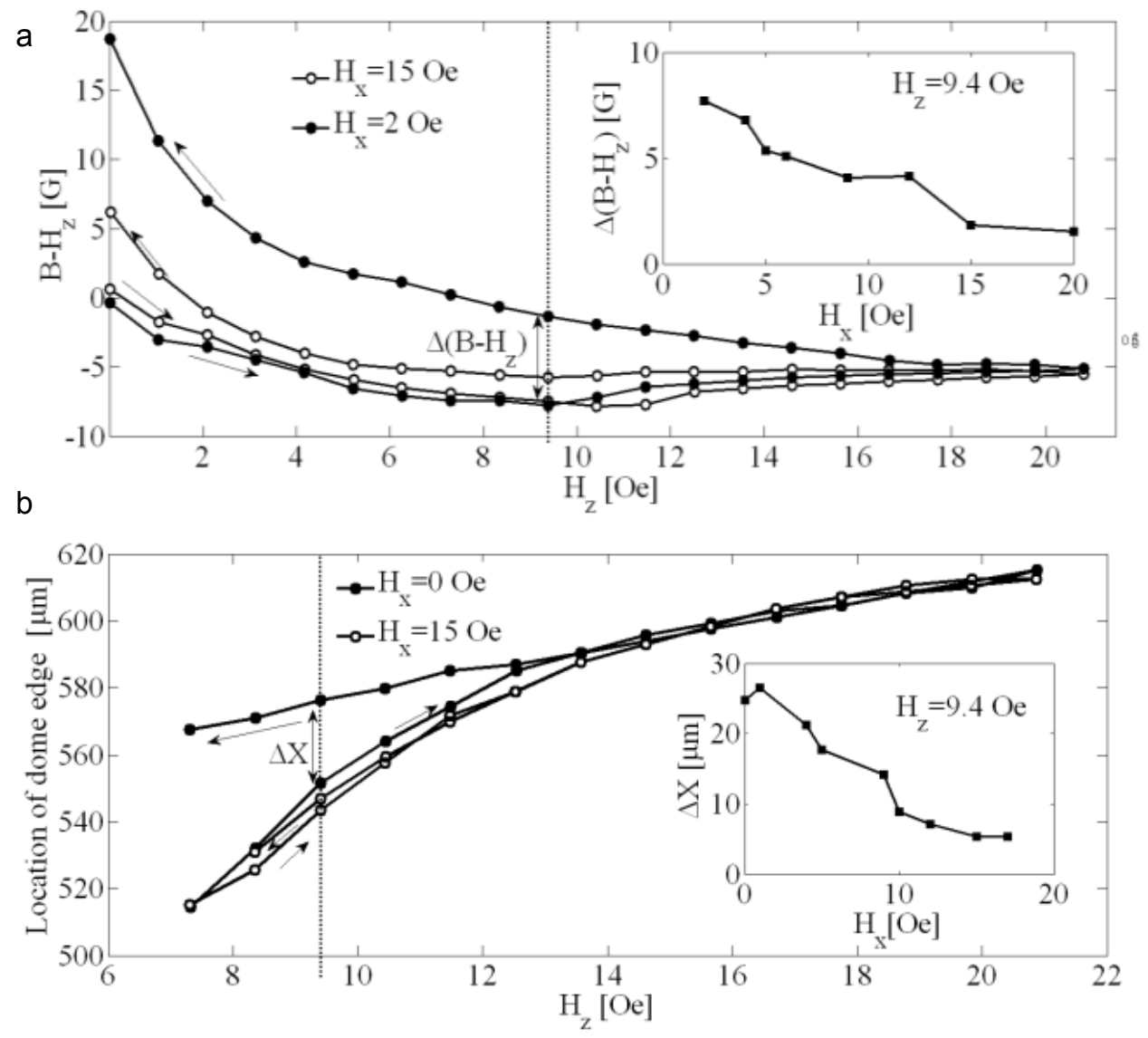

Figure 3: Correlation between two kinds of hysteresis is shown: a) hysteresis in local magnetization determined by B-$H_z$ dome edges location b) in dome edges location.

Fig. 3a shows magnetization loops for $H_x$=2 Oe (•) and 15 Oe (o). The dashed vertical line shows where the hysteresis was evaluated. In the inset, the width of the hysteresis was plotted at $H_z$=9.4 Oe. For $H_x$=15 Oe the width of the hysteresis is $\Delta$(B-Hz)=1.5 G while for $H_x$=2 Oe it is $\Delta$(B-Hz)=7.5 G, showing a hysteresis reduction by a factor of five from $H_x$=2 Oe to $H_x$=15 Oe.

Figure 3b shows the location of the dome edge while increasing and decreasing $H_z$ for $H_x$ = 0 (•) and 17 Oe (o). For the nonzero $H_x$, the vortex dome edge position becomes more reversible as the separation between the curves on increasing and decreasing $H_z$ diminishes. Clear vortex dome hysteresis suppression by $H_x$ can be seen in the inset for the same field $H_z$=9.4 Oe. For $H_x$=0 Oe the hysteresis width is $\Delta$X=25 μm, while for $H_x$=15 Oe it is $\Delta$X=5 μm.

Similar to Fig. 3a, here to, the hysteresis is reduced by a factor of five from $H_x$=0 Oe till $H_x$=20 Oe suggesting that one hysteresis behavior leads to the other. The hysteresis level ceases to decrease at the same equilibration field $H_x$=15 Oe. The functional dependence in the figures’ insets also look quite similar, pointing to a strong correlation between the hysteresis suppressions. Another similarity that can be noticed is that in both plots, the main contribution to the hysteretic behavior is on the decreasing field $H_z$.

**Due to this correlation, we conclude that for $H_z$ <20 Oe and T=82 K, $H_x$ affects hysteresis in local magnetization mainly through the geometrical barrier.**

## IV. DISCUSSION

To investigate the effect of $H_x$ on the geometrical barrier, we have plotted the dome width (the difference between the locations of the dome edges) as a function of $H_x$. Figure 4a is plotted for $H_z$=7 Oe on the decreasing field. The vortex dome’s width is reduced by 35% from $W_{dome}$=440 μm to $W_{dome}$=290 μm while $H_x$ is increased from 0 Oe to the equilibration field of |Hx|=15 Oe.

**We propose that suppression of the GB and equilibration of the vortex system are carried out by vortex chains, which act as channels connecting the vortex dome with the edges of the sample.**

To explain the strong effect of $H_x$ that causes the dome edge to move by over 35%, we made an energy estimation of a PV inside the dome and compared it with the crossing energy of a PV coupled by JVs outside the dome.

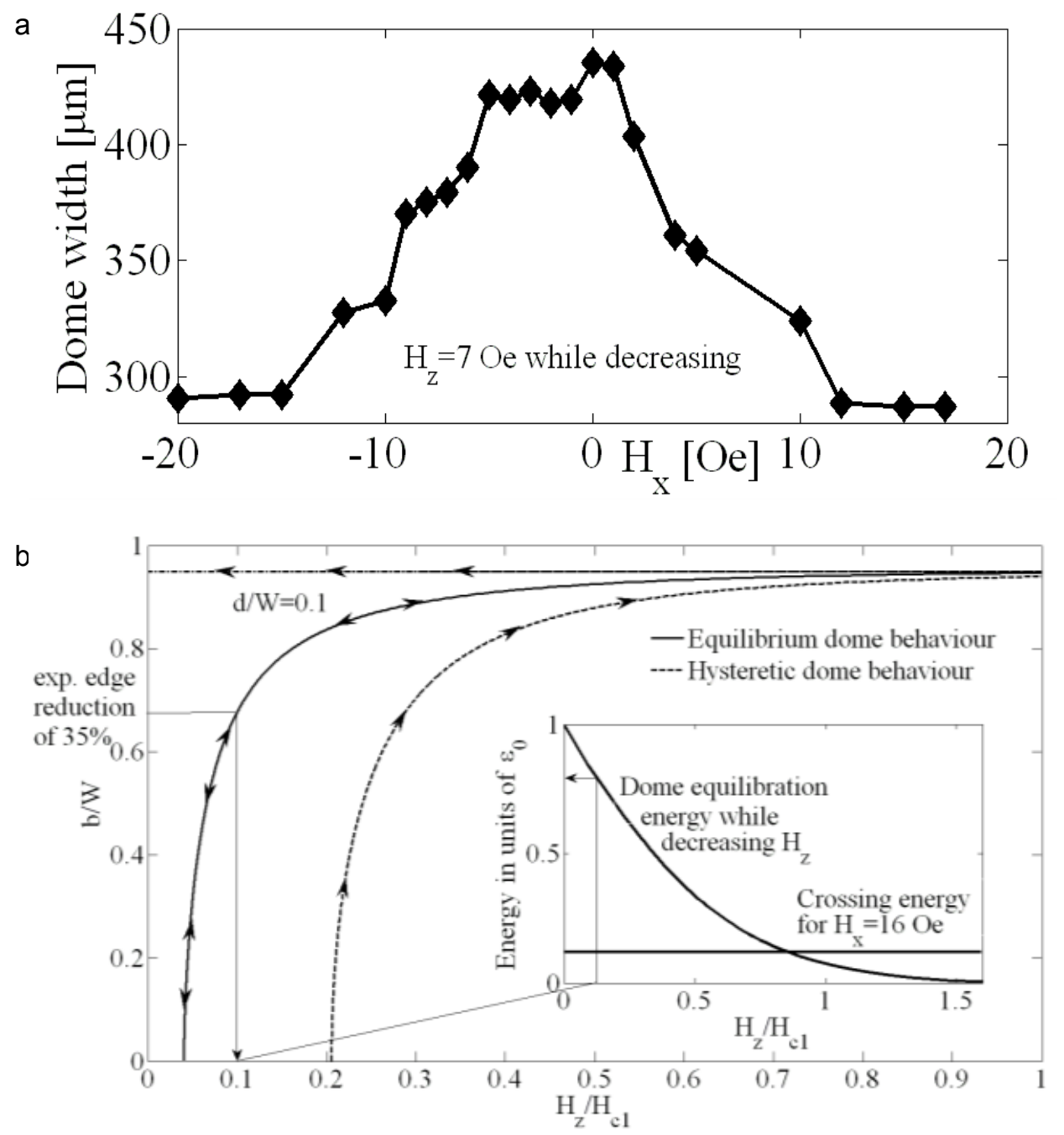

Figure 4: a) Dome width reduction due to applying in-plane field b) Equilibrium and hysteretic dependence of the dome edge on $H_z$. Inset: Calculation of the needed energy to equilibrate the dome and comparing it to the chains energy.

The energy that is needed to add a single vortex into the dome from outside the sample is obtained by integration of the Meissner current outside the dome [5]

(2) $$\varepsilon(x) = \varepsilon_0 (1 + \frac{4\pi}{H_{c1}} \int_x^W J_y(t) dt)$$

where $\varepsilon_0$ is the vortex line energy and W is the sample half width.
Using the current from the GB model [5], yields the following result:

(3) $$\varepsilon(x) = \varepsilon_0 \{1 - \frac{2}{d}[W - e + \frac{2}{\pi} W * I(x,b,e)]\}$$

where d is sample thickness, b and e are two parameters from GB model: half width of the dome and the location where vortices cut through the sample rims and

(4) $$I(x,b,e) = \int_{x/W}^{e/W} \arctan \sqrt{\frac{(1-(\frac{e}{W})^2)(t^2 - (\frac{b}{W})^2)}{(1-(\frac{b}{W})^2)((\frac{e}{W})^2 - t^2)}} dt$$

When this energy is zero, it will provide us the equilibrium dome edge behavior (continuous curve in Fig. 4b). This result is calculated and presented here for the first time. The dashed curves are the calculated dome edge behavior for increasing and decreasing $H_z$. A good agreement to the experimental data in Fig. 3b is seen.

The needed energy to equilibrate the dome while decreasing $H_z$ is calculated from Eq. 3 by setting b=W-d/2 and is plotted in the inset of Fig. 4b. Experimentally, we have seen in Fig. 4a that $H_x$ is efficient in equilibrating the vortex dome by up to 35% of the sample width. The Fig. 4b inset shows that for this equilibration energy of $0.8\varepsilon_0$ is needed.

The crossing energy between a single JV and a single PV, according to A. Koshelev [12], is given by

(5) $$E_x = -\frac{2.1\phi_0^2}{4\pi^2\gamma^2 s \cdot \ln(3.5\gamma s / \lambda_{ab})}$$

where $\lambda_{ab}$ is the penetration depth, s is the $CuO_2$ planes separation and $\gamma$ is the anisotropy parameter. The total number of JV layers in a unit thickness of a sample is

(6) $$N(H_x) = \frac{1}{d_c} = \sqrt{\frac{\sqrt{3}\gamma H_x}{2\phi_0}}$$

Thus, the total interaction energy is the sum of interactions between PV and all the JVs across the sample thickness. This is given by

(7) $$E_{xt}(H_x) = N(H_x) \cdot E_x$$

From previous calculations $\gamma$=406, $H_{c1}$=40 Oe and $\lambda$(T=82 K)=0.5 μm. By using a previous experimental estimation of the crossing energy by T. Tamegai [13] $E_x(theor.)/E_x(\exp.) = 0.15$ we obtain the following

(8) $$E_{xt}(H_x) = -\sqrt{H_x}\varepsilon_0 / 39$$ and

(9) $$E_{xt}(H_x = 15Oe) \approx -0.1\varepsilon_0 .$$

## V. CONCLUSION

We have shown that the vortex dome energy and the crossing energy scales are of the same order of magnitude but have a difference of a factor of 8 which needs to be investigated theoretically. **Quantitative energy estimations show that the vortex chain energy is high enough to move the dome edges, resulting in the whole system's equilibration.**

**Email: Ilia.Gutman@weizmann.ac.il*